\documentclass[letterpaper, 10 pt, conference]{ieeeconf}  

\IEEEoverridecommandlockouts                              

\usepackage{graphicx} 
\usepackage{subfigure}
\usepackage{mathptmx} 
\usepackage{times} 
\usepackage{amsmath} 
\usepackage{amssymb}  
\usepackage{mathtools} 
\usepackage{enumerate} 
\usepackage{booktabs}
\usepackage{soul}
\usepackage{color}
\usepackage{cite}
\usepackage{ragged2e}
\usepackage{mathrsfs}

\newcounter{theorem}
\newtheorem{proposition}[theorem]{Proposition}
\newtheorem{remark}[theorem]{Remark}
\newcommand{\T}{^{\mathrm{T}}}							  
\newcommand{\B}[1]{\if#1\relax\bm{#1}\else\mathbf{#1}\fi} 
\newcommand{\R}[1]{\mathrm{#1}}						      

\makeatletter
\def\amsbb{\use@mathgroup \M@U \symAMSb}
\makeatother

\title{\LARGE \bf
Adaptive and quasi-sliding control of shimmy in landing gears
}

\author{Daniel A. Burbano-L$^{1}$, Marco Coraggio$^{2}$, Mario di Bernardo$^{2}$, Franco Garofalo$^{2}$ and Michele Pugliese$^{2}$
\thanks{*This work was developed within the frame of the Project CAPRI- Landing Gear System with intelligent Actuation, co-financed by MIUR-Italian Ministry of Research with DAC-Campania Aerospace District as beneficiary and Magnaghi Aeronautica SpA as ``prime'' partner.}
\thanks{$^{1}$ Department of Electrical Engineering and Computer Science, Northwestern University, Evanston, IL 60208-3118, USA
        {\tt\small danielburbano@northwestern.edu}.}%
\thanks{$^{2}$ Department of Electrical Engineering and Information Technology, University of Naples Federico II, Via Claudio 21, 80125 Naples, Italy
        {\tt\small \{marco.coraggio, mario.dibernardo, mipuglie,}
        {\tt\small franco.garofalo\}@unina.it}.}%
}

\begin{document}

\maketitle
\thispagestyle{empty}
\pagestyle{empty}

\begin{abstract}
Shimmy is a dangerous phenomenon that occurs when aircraft's nose landing gears oscillate in a rapid and uncontrollable fashion.
In this paper, we propose the use of two nonlinear control approaches (zero average control and model reference adaptive control based on minimal control synthesis) as simple yet effective strategies to suppress undesired oscillations, even in the presence of uncertainties and partial state measurements. Numerical results are presented to validate the proposed control approaches. 
\end{abstract}
%
%
\section{Introduction}
When taking-off, landing or taxiing, any airplane might experience vibrations due to unstable oscillation of the nose landing gear (NLG).
This phenomenon, also known as \emph{shimmy}, is often unpredictable with consequences ranging from annoying vibrations to serious damage or even collapse of the airplane \cite{besselink2000}.
Shimmy is not an exclusive problem in aeronautics; in fact, motorcycles and cars also display a similar issue which is often termed as \emph{wobble} \cite{limebeer2006}. 

The study of shimmy can be traced back to the early 20's when the first tires were manufactured \cite{ran2016}.
Since then, many mathematical models have been proposed aiming at replicating these oscillations and uncovering the main causes for their emergence (see \cite{besselink2000,takacs2010,ran2016} and references therein for a detailed list of shimmy models). 
One of the main causes of shimmy is the interaction of the tire with the road.
As a matter of fact, the presence of a lateral force on the tire produces a side slip angle, and thus rotations of the NLG. 
This in turn produces further lateral forces on the tire, thus creating a repetitive (positive-feedback) loop that causes oscillations. 
Recently, it has also been shown that shimmy can be caused by other nonlinear effects such as friction \cite{zhuravlev2013}, free-play, and gyroscopic forces \cite{howcroft2015}.  
Therefore, designing and implementing control approaches for suppressing shimmy despite model uncertainties is of great importance for reliability and safety of airplanes on the ground. 
It is worth mentioning that shimmy control strategies should guarantee fast convergence and most importantly a small overshoot of approximately one degree at most. 
The classic solutions to reduce shimmy are based on the adoption of passive strategies, where the aim is to increase the NLG stiffness and damping constant by using different construction materials or additional passive dampers, respectively \cite{pouly2011}. 
However, airplanes are subject to a plethora of unexpected and dynamic disturbances like changing loads and nonuniform tire-road interfaces; in addition, aircrafts require frequent maintenance, making passive approaches less effective and costly \cite{tourajizadeh2016}. 

Moreover, recent developments in the aircraft industry are aimed at implementing fly-by-wire strategy, replacing mechanical and pneumatic actuators by electromechanical devices \cite{boglietti2009}.
In fact, novel steering architectures consider electromechanical actuators where the gear rotation is controlled by a brush-less motor located on the top of the turning tube \cite{liscouet2012}.
Within this context, the use of active control solutions for suppressing shimmy can be easily implemented \cite{bonfe2011}.
Indeed, in the last decade, different active control strategies for shimmy have been proposed.
For instance, in \cite{goodwine2000}, a feedback linearization approach based on full state measurements has been shown to be effective in suppressing the oscillations; however, nonlinearities in the model are assumed to be perfectly known.
This approach was later extended in \cite{pouly2011} to the case where the nonlinear functions are unknown and they are estimated using adaptive strategies and fuzzy logic theory.
More recently, in \cite{hajiloo2015}, a robust model predictive control was designed for a linearized model of an NLG.
Differently, in \cite{tourajizadeh2016}, a nonlinear optimal control was presented based on the use of state-dependent Riccati equations; furthermore, a switching action in the controller was added to better stabilize the closed-loop system.
Finally, a simpler PID controller is utilized in \cite{orlando2017}, where the control gains are designed using a decline population swarm optimization technique.

Previous control approaches either consider linearized models and assume full state measurements to be available or adopt very complex and sophisticated control solutions \cite{pouly2011,hajiloo2015,tourajizadeh2016,orlando2017}. 
In contrast, in this paper we only use partial state measurements and we also test the controllers in the case that the nonlinear terms in the model are uncertain.
In particular, we propose the use of either a model reference adaptive control (MRAC) with minimal control synthesis (MCS) \cite{stoten1990}, or a zero-average dynamics (ZAD) control \cite{fosas2000}.
To reconstruct the inaccessible states, and hence close the loop, we make use of a classic Luenberger observer which is designed under the assumption that the vector field is QUAD \cite{delellis2011} rather than Lipschitz continuous (as is usually done in this type of problem).
We show that the QUAD condition provides less conservative results, so that lower values of the control gains can be chosen, thus avoiding large overshoots in the closed-loop system.

%
\section{Problem Statement}
\label{sec:Problem Statement}
\subsection{Nose landing gear model}
\begin{figure}[tbp]
\centering {
\subfigure[]
{\label{fig:1:a}
{\includegraphics[scale=0.17]{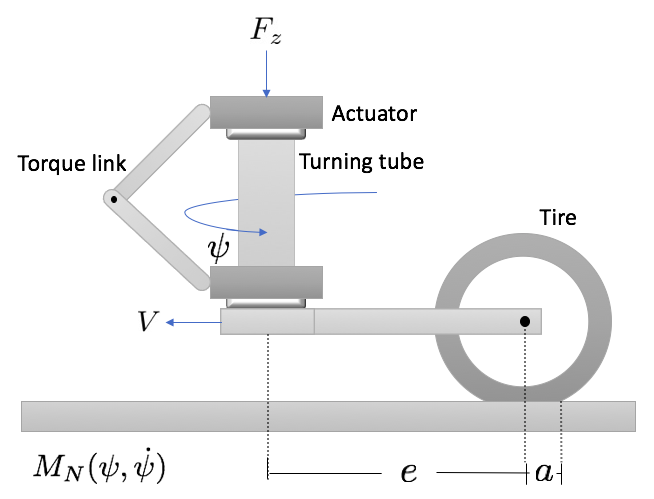}}}
\subfigure[]
{\label{fig:1:b}
{\includegraphics[scale=0.17]{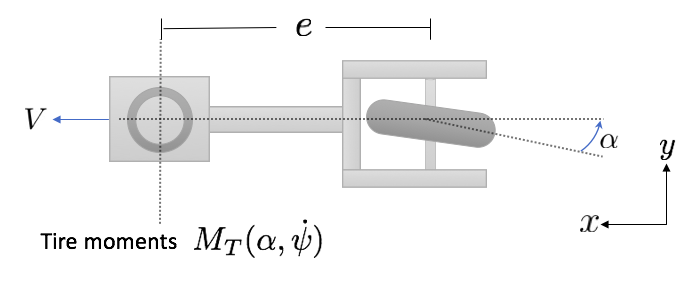}}}
}
\caption{Simplified model of an NLG: (a) side and (b) top views.}
\label{fig:1}
\end{figure}

\begin{figure}[t]
\centering {
\subfigure[]
{\label{fig:2:a}
{\includegraphics[scale=0.2]{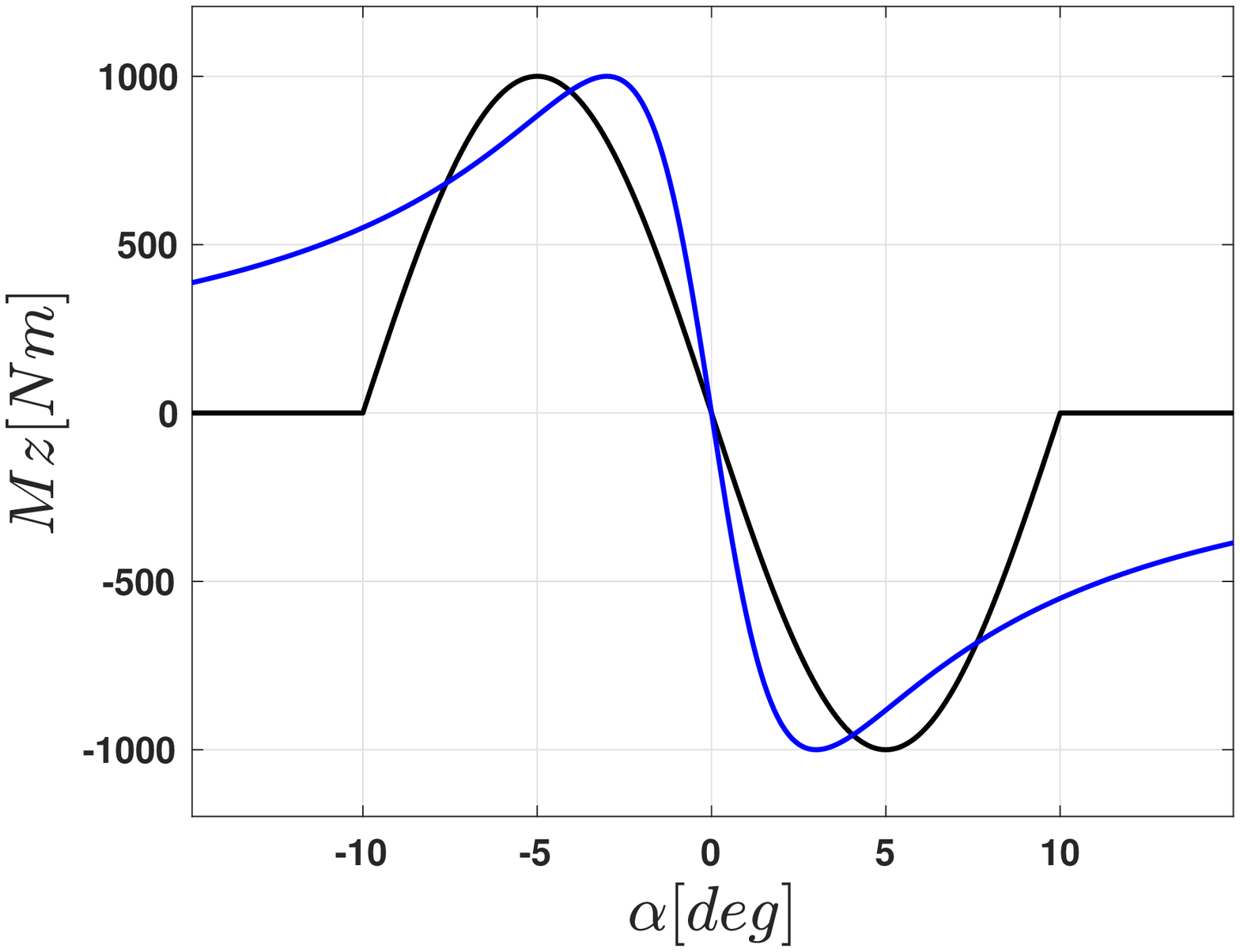}}}
\subfigure[]
{\label{fig:2:b}
{\includegraphics[scale=0.2]{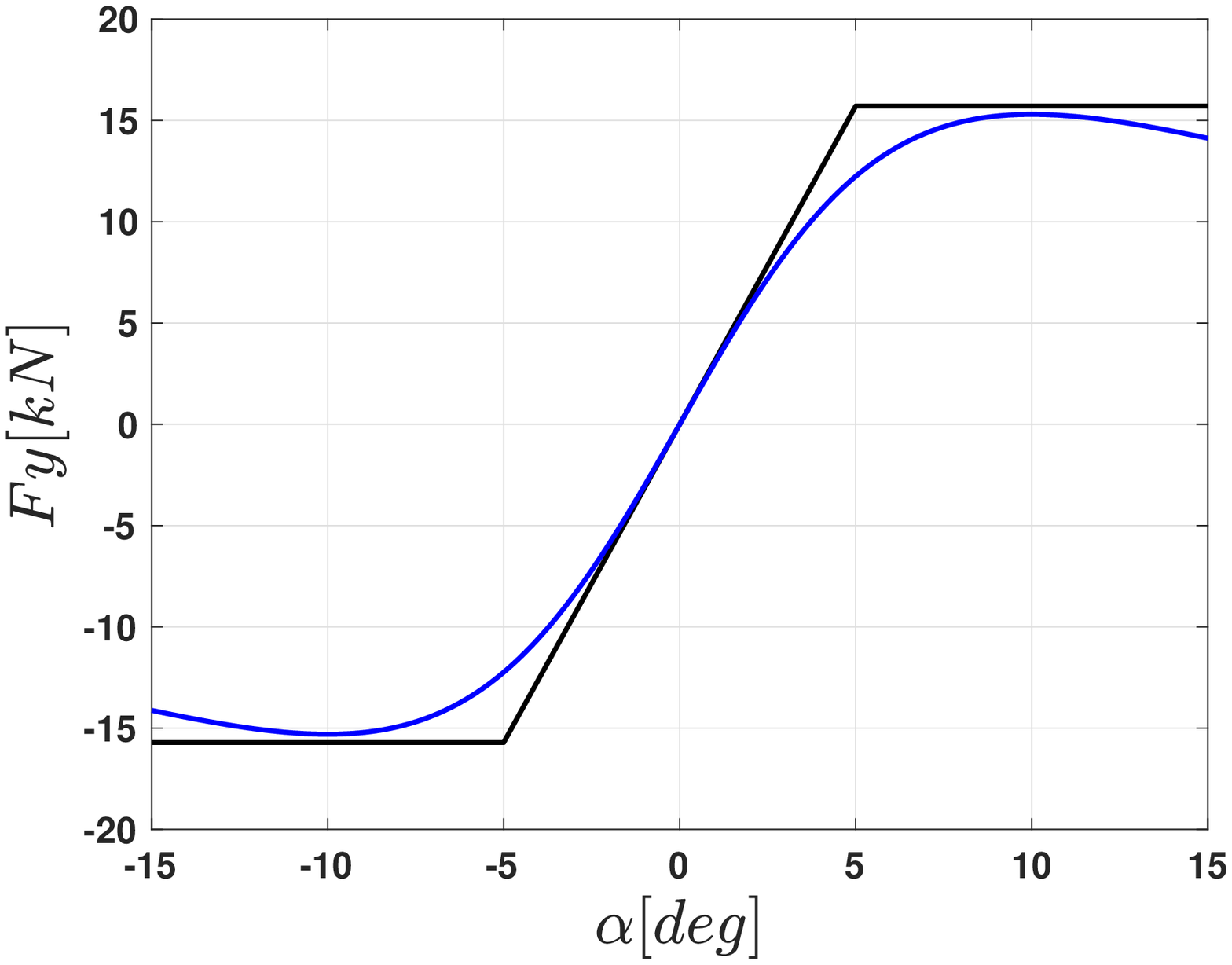}}}
}
\caption{Approximating functions for $M_z(\alpha)$ (a) and $F_y(\alpha)$ (b). Black and blue lines represent the piece-wise and smooth approximations, respectively.}
\label{fig:2}
\end{figure}

We consider a simplified model of a single wheeled aircraft's nose landing gear (NLG) with an electromechanical actuator (see Figure \ref{fig:1}), which is described by the third order nonlinear differential equation given by \cite{somieski1997}
\begin{align}
\label{Eq:1}
I_z\ddot{\psi}(t) &= M_{\R{N}}(\psi(t),\dot{\psi}(t)) + M_{\R{T}}(\alpha(t),\dot{\psi}(t))+\tau(t),\\
\sigma\dot{\alpha}(t) &=  V(\psi(t)-\alpha(t)) + (e-a)\dot{\psi}(t),
\end{align}
where $\psi(t) \ [\mathrm{rad}]$ and $\alpha(t) \ [\mathrm{rad}]$ are the state variables denoting the yaw and slip angles respectively, $I_z$ is the moment of inertia about the $z$-axis, $V$ is the wheel forward velocity, $e$ is the wheel caster length, $a$ is the half contact length of the tire on the ground, and $\sigma$ is the relaxation length of tire deflection.
The torque $M_{\R{N}}(\psi(t),\dot{\psi}(t)):=c\psi(t)+k\dot{\psi}(t)$ is the sum of a linear elastic torque provided by the turning tube, with constant torsional rate $c$, and a linear damping term, with coefficient $k$, that models viscous frictions coming from the shock absorber on the bearing of the oil-pneumatic and the shimmy damper.
An external torque, $\tau(t):=u(t) + \zeta(t)$, models the action $u(t)$ exerted by the control input and some disturbance $\zeta(t)$.  
For what concerns the values of the parameters, we use those reported in  Table \ref{Table:1}, as in \cite{somieski1997}.
\begin{table}[t]
\caption{NLG parameters}
\label{Table:1}
\begin{center}
\begin{tabular}{@{}llll@{}}
\toprule
Parameter                   &  Symbol       &  Value    &  Unit       \\
\midrule
velocity                    & $V$           & $[0,80]$  & m/s        \\
half contact length         & $a$           & $0.1$     & m          \\
caster length               & $e$           & $0.1$     & m          \\
moment of inertia           & $I_z$         & $1$       & kg$\cdot$m\textsuperscript{2}   \\
vertical force              & $F_z$         & $9000$    & N          \\
torsional spring rate       & $c$           & $-100000$ & N$\cdot$m/rad     \\
side force derivative       & $ c_{F,\alpha}$ & $20$      & 1/rad      \\
moment derivative           & $ c_{M,\alpha}$ & $-2$      & m/rad      \\
torsional damping constant  & $k$           & $-10$     & N$\cdot$m/rad/s   \\
tread width moment constant & $\kappa$      & $-270$    & N$\cdot$m\textsuperscript{2}/rad \\
relaxation length           & $\sigma=3a$   & $0.3$     & m          \\
\bottomrule
\end{tabular}
\end{center}
\end{table}
Moreover, $M_{\R{T}}(\alpha(t),\dot{\psi}(t)):= M_{\R{D}}(\dot{\psi}) + M_{\R{G}}(\alpha)$ represents the tire moments originated from tire damping and deformations.
Specifically, $M_{\R{D}}(\dot{\psi}):=(\kappa/V) \psi(t)$ is the damping term with coefficient $\kappa$, whereas $M_{\R{G}}(\alpha)$ describes the interaction between the tire and the ground.
This interaction is highly nonlinear and is due to lateral tire deformations caused by side slip; it is given by
\begin{equation}
    M_{\R{G}}(\alpha)=M_z(\alpha) - eF_y(\alpha),
\end{equation}
where $M_z(\alpha)$ and $F_y(\alpha)$ are nonlinear functions representing the aligning torque about the tire's center and the tire side force, respectively.
In particular, we consider two different pairs of functions approximating $M_z(\alpha)$ and $F_y(\alpha)$.
The first one is a 
\textit{piece-wise smooth approximation}:
\begin{align}
M_{z,1}(\alpha) &= 
\begin{dcases}
{F_z c_{M,\alpha}\frac{\alpha_{\R{g}}}{\pi}\sin \left( \frac{\pi\alpha}{\alpha_{\R{g}}} \right)}, & {\textup{if } \vert \alpha \vert\le \alpha_{\R{g}}} \\
{0}, & \textup{otherwise}
\end{dcases},\label{eq:M_nonsmooth}\\
F_{y,1}(\alpha) &= \frac{ c_{F,\alpha} F_z}{2}\left( \vert \alpha+\delta \vert - \vert \alpha-\delta \vert\right),
\label{eq:F_nonsmooth}
\end{align}
where $\alpha_{\R{g}}=10\pi/180$ and $\delta = 5\pi/180$.
The second one is a \textit{smooth approximation}:

\begin{align}
  M_{z,2}(\alpha) &=  c_{M,\alpha}F_z\gamma_M\frac{2\alpha\alpha_M}{\alpha^2+\alpha_M^2},\label{eq:M_smooth}\\
  F_{y,2}(\alpha) &=  c_{F,\alpha}F_z\gamma_F\frac{2\alpha\alpha_F}{\alpha^2+\alpha_F^2},
  \label{eq:F_smooth}
\end{align}
where $\alpha_M=3\pi/180$, $\gamma_M=0.1\alpha_{\R{g}}/\pi$, $\alpha_F=3\alpha_{\R{g}}$, and $\gamma_F=0.085$. 
The functions $M_{z,1}(\alpha)$, $M_{z,2}(\alpha)$, $F_{y,1}(\alpha)$, and $F_{y,2}(\alpha)$ are shown in Figure \ref{fig:2}. 
%
%
%

The NLG dynamics can be written in compact form as
\begin{equation}
\begin{aligned}
&\dot{\mathbf{x}}(t)=\mathbf{A}\mathbf{x}(t) + \mathbf{f}(\mathbf{x}) + \mathbf{B}u(t) + \mathbf{B}\zeta(t), \quad \mathbf{x}(0)=\mathbf{x}_{0}, \\
&\mathbf{y}(t) = \mathbf{C} \mathbf{x}(t), 
\quad \mathbf{C} = \left[ {\begin{array}{ccc}
   {1} & 0 & 0\\
   {0} & {1} & 0
  \end{array} } \right],
\label{main:eq}
\end{aligned}
\end{equation}
where $\mathbf{x}:= \begin{bmatrix} \psi & \dot{\psi} & \alpha \end{bmatrix}\T$, $\mathbf{y}$ is the output, $\mathbf{f}(\mathbf{x}):= \begin{bmatrix} 0 & M_{\R{G}}(\alpha)/I_z & 0 \end{bmatrix}\T$, $\mathbf{B}:= \begin{bmatrix} 0 & 1/I_z & 0 \end{bmatrix}\T$, and 
\begin{equation}
\mathbf{A}:=  \left[ {\begin{array}{ccc}
   {0} & 1 & 0\\
   {c/I_z} & (k/I_z)+(\kappa/(I_zV)) & 0\\
   V/\sigma & (e-a)/\sigma & -V/\sigma
  \end{array} } \right].
\end{equation}
Note that we only consider the yaw angle $\psi$ and its velocity $\dot{\psi}$ as the available outputs.
In fact, $\psi$ can be measured using a RVDT (Rotary Variable Differential Transformer) sensor on the NLG, while the velocity $\dot{\psi}$ can be easily obtained from $\psi$ or using a dedicated sensor \cite{pouly2011}.
Moreover, the state variable $\alpha$ is related to the lateral displacement of the tire, and, as a consequence, it is much more cumbersome to design appropriate sensors to accurately measure this variable \cite{pouly2011}. 
Therefore, the challenge is to design an appropriate observer for reconstructing this missing state, together with robust controllers that suppress undesired oscillations, while guaranteeing fast convergence and overshoots lower than 1\textsuperscript{$\circ$}.
\subsection{Open-loop dynamics: bifurcation diagrams}
\begin{figure}[t]
\centering {
\subfigure[]
{\label{fig:3:a}
{\includegraphics[scale=0.2]{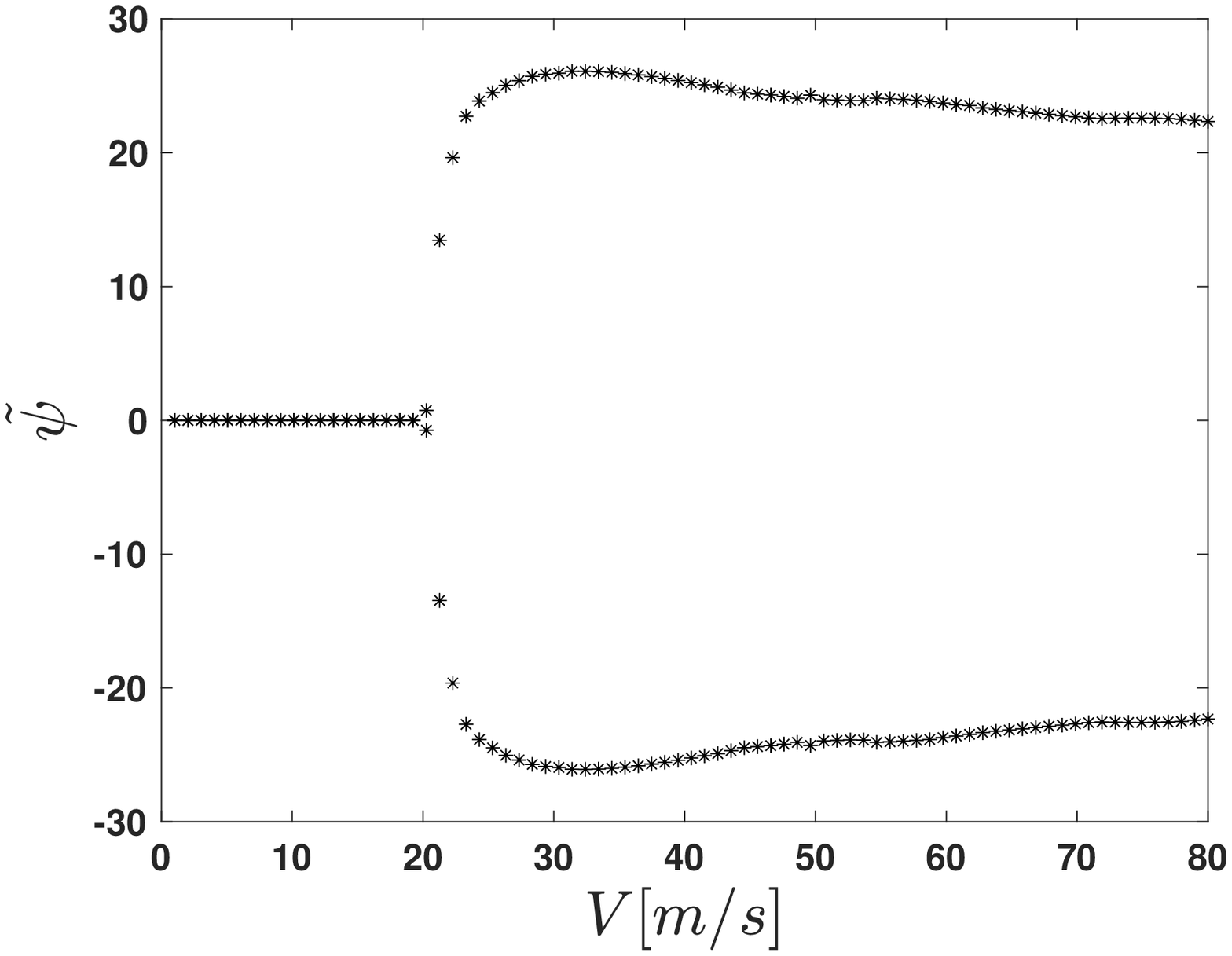}}}
\subfigure[]
{\label{fig:3:b}
{\includegraphics[scale=0.2]{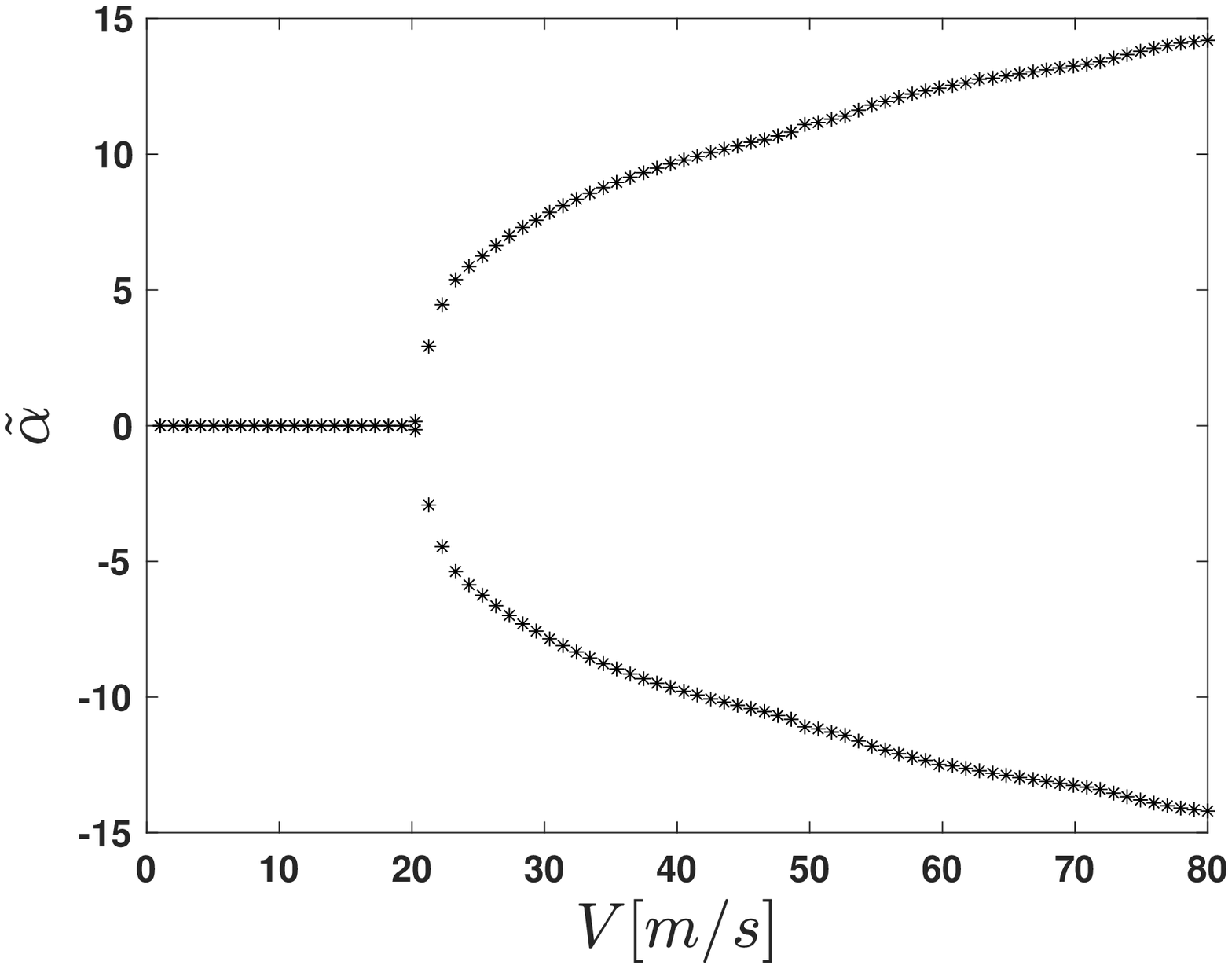}}}
\subfigure[]
{\label{fig:3:c}
{\includegraphics[scale=0.2]{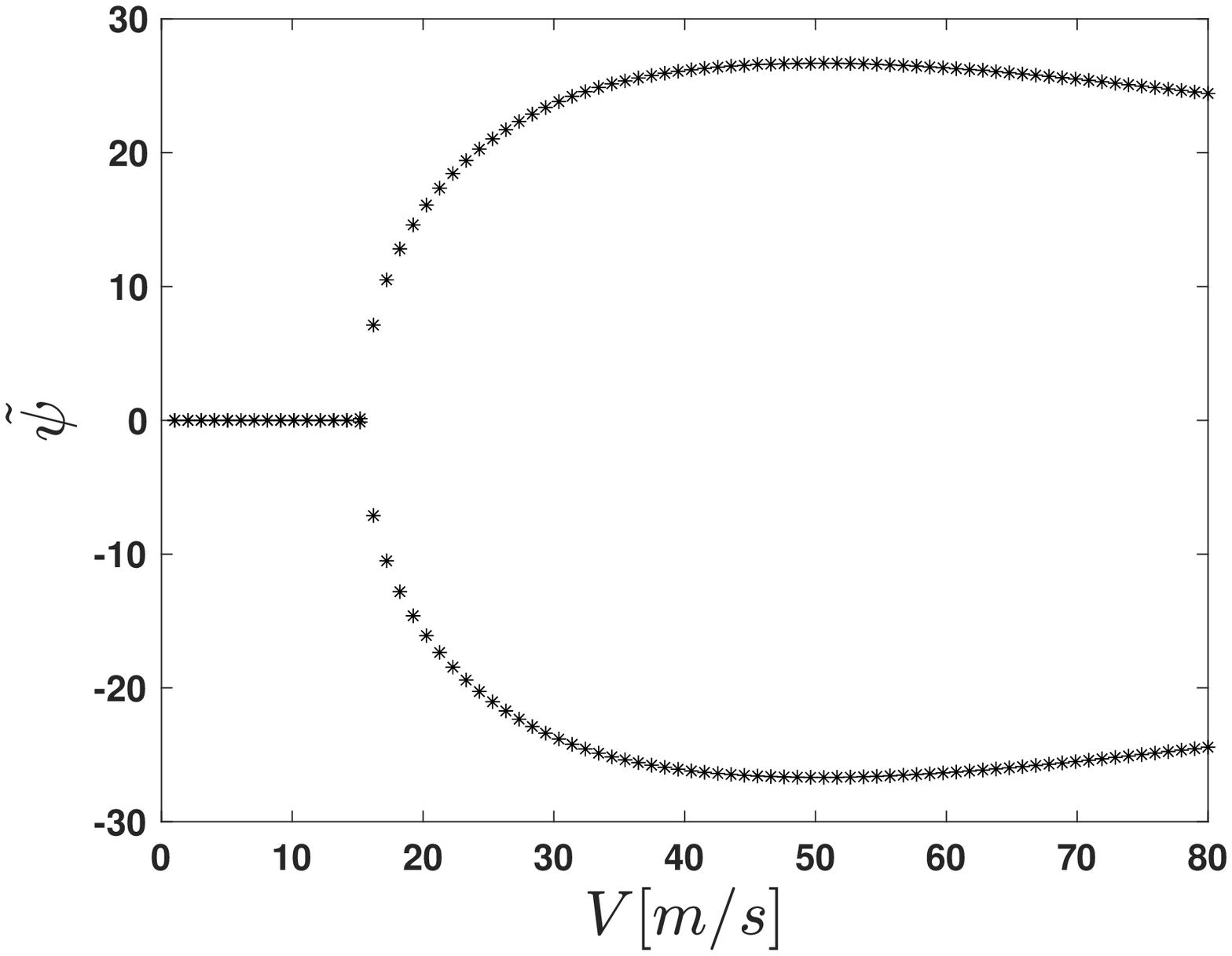}}}
\subfigure[]
{\label{fig:3:d}
{\includegraphics[scale=0.2]{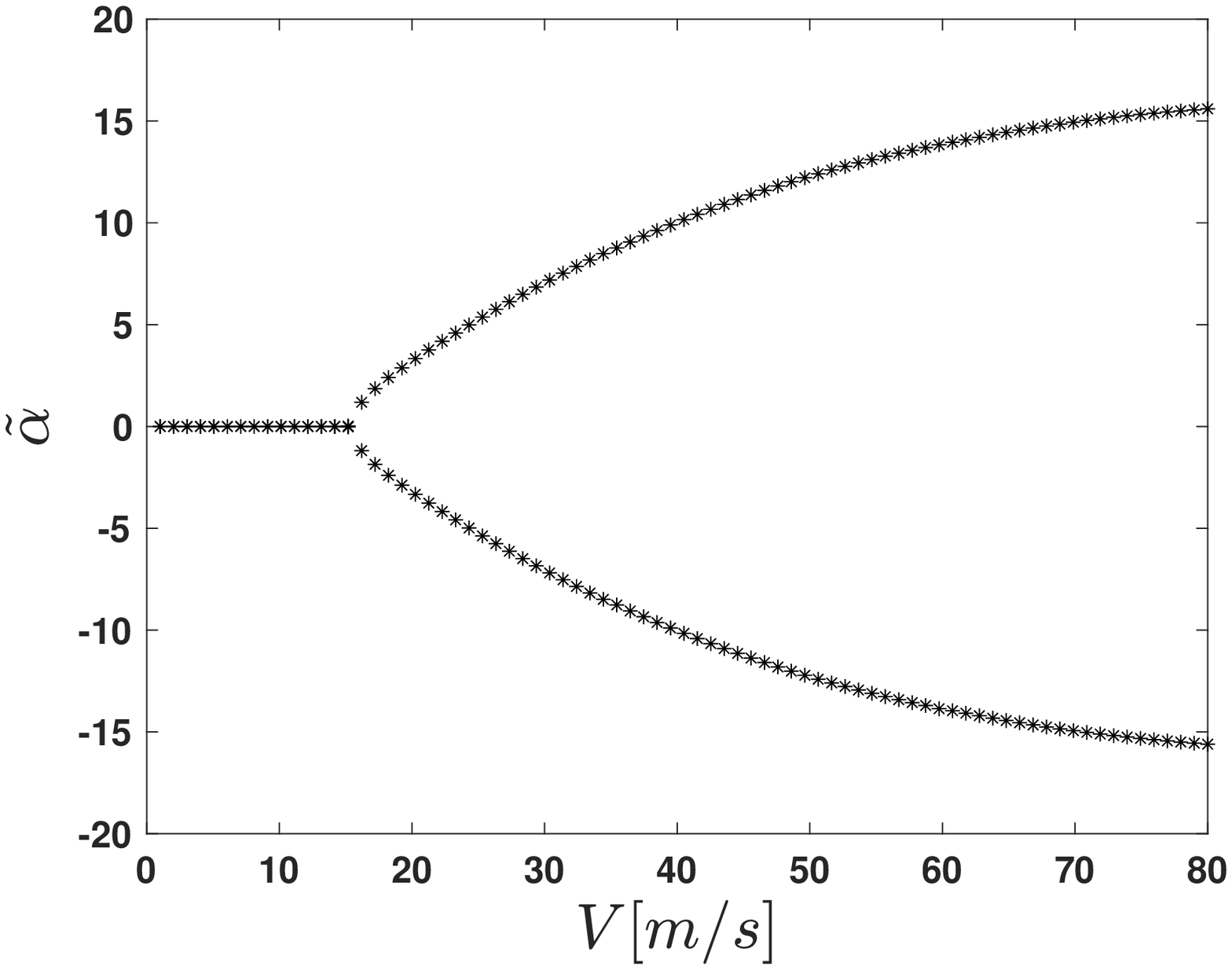}}}
}
\caption{Bifurcation diagrams varying the velocity $V$ with the nonlinear function $\mathbf{f}(\mathbf{x})$ being: (a, b) non-smooth, (c, d) smooth.
$\widetilde{\psi}$ (a, c) and $\widetilde{\alpha}$ (b, d) represent the minimum and maximum amplitude of the steady state trajectory for $\psi$ and $\alpha$, respectively.}
\label{fig:3}
\end{figure}

To illustrate the control problem, we start by showing the NLG dynamics in the absence of control ($u(t) = 0$). 
More specifically, we describe how using either of the two different approximations for the nonlinear forces $M_z$ and $M_y$ affects the system dynamics.
To that aim, we compute bifurcation diagrams of system \eqref{main:eq} for both approximations.
Indeed, bifurcation diagrams are an important tool for analysis of nonlinear systems and have been recently used for studying shimmy behavior \cite{howcroft2015}.
We select the forward velocity $V$ as bifurcation parameter, as increasing or decreasing it corresponds to the common scenarios of taking off, landing or taxiing.
We choose to vary the velocity in the interval $[0,80]$ m/s---common in small planes---using unitary steps.
For each value of the forward velocity, we plot the minimum and maximum amplitudes $\widetilde{\psi}$ and $\widetilde{\alpha}$ of the steady state response for the states $\psi$ and $\alpha$, respectively (see Figure \ref{fig:3}).
Note that, for both the piece-wise smooth and the smooth approximation, we observe the system undergo a Hopf bifurcation at $V=20$ and $V=16.5$, respectively.
This suggests that different approximations of the nonlinear moment $M_{\R{D}}$ shift the bifurcation point. 
Hence, designing robust control strategies that can cope with uncertainty on the nonlinear forces $M_z$ and $M_y$ is crucial, given that in real world scenarios these moments are not known exactly or their parameters might change over time.
Moreover, the peak oscillation amplitude $\widetilde{\psi}$ is found to be 26.4 for both approximations, and is reached when $V=30$ and $V=46.5$, respectively.
As for the frequency of the oscillations, in both cases it is approximately 50 Hz.
\section{Controlling Shimmy}
\label{sec:Controlling Shimmy}
We employ two different control strategies for attenuating shimmy vibrations: namely, a Zero Average Dynamics (ZAD) control \cite{fosas2000} and a model reference control based on Minimal Control Synthesis (MCS) \cite{stoten1990}.
These control strategies require full knowledge of the state variables; however, we only have access to the output $\mathbf{y}(t)$.
Hence, first we need to design an observer for reconstructing the missing states.
\subsection{Observer design}
Consider the classic Luenberger observer given by
\begin{equation}
\dot{\hat{\mathbf{x}}}(t) = \mathbf{A}\hat{\mathbf{x}}(t) + \mathbf{B}u(t) + \mathbf{f}(\hat{\mathbf{x}}(t)) - \mathbf{L}(\mathbf{C}\hat{\mathbf{x}}(t)-\mathbf{y}(t)).
\label{eq:obs}
\end{equation}
The pressing challenge is to design appropriately the matrix $\mathbf{L}$ that guarantees $\hat{\mathbf{x}}(t) \rightarrow \mathbf{x}(t)$ as $t\rightarrow\infty$.
To that aim we define the observation error $\mathbf{e}(t):=\hat{\mathbf{x}}(t)-\mathbf{x}(t)$ and present the following result.
\begin{proposition} Observer \eqref{eq:obs} asymptotically reconstructs the states of system \eqref{main:eq}, that is, $\lim_{t \rightarrow \infty} \Vert\mathbf{e}(t)\Vert=0$, if the following conditions are fulfilled:
\begin{enumerate}[(i)]
\item there exist a nonzero constant $\rho$ and a symmetric positive definite matrix $\mathbf{P}$ such that $(\mathbf{v}_1-\mathbf{v}_2)\T\mathbf{P}(\mathbf{f}(\mathbf{v}_1)-\mathbf{f}(\mathbf{v}_2))\le\rho (\mathbf{v}_1 - \mathbf{v}_2)\T(\mathbf{v}_1-\mathbf{v}_2), \forall \ \mathbf{v}_1,\mathbf{v}_2 \in \Omega\subseteq\amsbb{R}^{3}$,
\item there exist a generic matrix $\mathbf{L}$ and a symmetric positive definite matrix $\mathbf{Q}$ such that $\mathbf{P}(\mathbf{A}-\mathbf{L}\mathbf{C})+(\mathbf{A}-\mathbf{L}\mathbf{C})\T\mathbf{P}=-\mathbf{Q}$,
\item $\lambda_{\mathrm{min}}(\mathbf{Q}) >\rho$.
\end{enumerate}
\label{propo:1}
\end{proposition}
\begin{proof} The proof follows from choosing $V=\mathbf{e}\T\mathbf{P}\mathbf{e}$ as candidate Lyapunov function \cite{rajamani1998}, yielding $\dot{V}\le -\mathbf{e}\T\mathbf{Q}\mathbf{e}+\mathbf{e}\T\mathbf{P}(\mathbf{f}(\hat{\mathbf{x}})-\mathbf{f}(\mathbf{x}))$.
Then, $\dot{V}< 0$ if the three conditions are fulfilled, which completes the proof.
\end{proof}

\begin{remark} 
We wish to highlight two facts.
Firstly, condition (i) is known as \emph{QUAD} (quadratic) condition and it has been widely used for proving convergence in complex networks \cite{delellis2011} (the special case when $\mathbf{P}=\mathbf{I}_n$ is also known as \emph{one-sided Lipschitz continuity} condition \cite{cortes2008,zhang2012}).
In fact, several nonlinear possibly chaotic systems satisfy this condition \cite{delellis2011}.
Moreover, condition (ii) is standard within the context of observer design \cite{rajamani1998,zhang2012}, and can be easily solved by using standard optimization software.
Secondly, one of the main issues when designing Luenberger observers for nonlinear systems are the restrictive synthesis conditions based on Lipschitz continuity of the vector-fields \cite{rajamani1998}.
In fact, for some systems, this condition might not be fulfilled or the Lipschitz constant might be excessively large, yielding to overly conservative results \cite{zemouche2013}.
Indeed, as we show below, the high value of the Lipschitz constant for the NLG model considered here yields a matrix $\mathbf{L}$ with large entries.
Although convergence is guaranteed, performance is not, since high gains (entries of $\mathbf{L}$) would cause overshoots larger than $1^\circ$.
On the other hand, the QUAD condition is more general than the Lipschitz condition \cite{delellis2011} (i.e.~a wider class of nonlinear systems satisfy it), and most importantly it provides less conservative results.
As a matter of fact, any contracting vector field or system with bounded Jacobian is QUAD \cite{delellis2011}.
\end{remark}

Next, we use Proposition \ref{propo:1} to find the matrix $\mathbf{L}$. 
Thus, we first start by noticing that the nonlinear function $\mathbf{f}(\mathbf{x})$ given by either \eqref{eq:M_nonsmooth}-\eqref{eq:F_nonsmooth} or \eqref{eq:M_smooth}-\eqref{eq:F_smooth} has bounded Jacobian.
For the sake of simplicity, we only consider the piece-wise linear function \eqref{eq:M_nonsmooth}-\eqref{eq:F_nonsmooth}, whose Jacobian matrix $\mathbf{Df}$ is given by
$$
\mathbf{Df} = \left[ {\begin{array}{ccc}
   {0} & {0} & {0}\\
   {0} & {0} & {Df_{23}}\\
   {0} & {0} & {0}
  \end{array} } \right],
$$
where
$$
Df_{23} = 
\begin{dcases}
{m_1 \cos( \alpha \pi / \alpha_{\R{g}} )-m_2}, & {\textup{if } \vert \alpha \vert\le \delta} \\
{m_1 \cos( \alpha \pi / \alpha_{\R{g}} )}, & {\textup{if } \vert \alpha \vert> \delta} \ \textup{and} \ \vert \alpha \vert\le \alpha_{\R{g}}\\
{0}, & {\textup{otherwise}}
\end{dcases},
$$
with $m_1 := F_z c_{M,\alpha}$ and $m_2 = e c_{F,\alpha}F_z$.
Next, we define $\mathbf{g}(\theta) := \mathbf{f}(\mathbf{v}_2 + \theta(\mathbf{v}_1 - \mathbf{v}_2))$, for $\theta \in [0,1]$. 
From the fundamental theorem of calculus, one has $\mathbf{f}(\mathbf{v_1})-\mathbf{f}(\mathbf{v_2})=\mathbf{g}(1)-\mathbf{g}(0) = 
\int_0^1{(\R{d}\mathbf{g}(\theta)/\R{d}\theta)\R{d}\theta} = 
[\int_0^1{\mathbf{Df}(\mathbf{v}_2+\theta(\mathbf{v}_1-\mathbf{v}_2)) \R{d}\theta}] (\mathbf{v}_1-\mathbf{v}_2)$.
Thus, we have that $\left\|\mathbf{f}(\mathbf{v}_1)-\mathbf{f}(\mathbf{v}_2)\right\|\le \sup_{\theta\in [0,1]} \left\|\mathbf{Df}(\mathbf{v}_2+\theta(\mathbf{v}_1-\mathbf{v}_2))\right\|\left\|\mathbf{v}_1-\mathbf{v}_2\right\|$, 
and from the fact that
$ \sup_{\theta\in [0,1]} \left\|\mathbf{Df}(\mathbf{v_2}+\theta(\mathbf{v}_1-\mathbf{v}_2))\right\|\le\vert Df_{23}\vert =\vert m_1 \cos(\alpha \pi / \alpha_{\R{g}} ) - m_2\vert$, we have
$\left\|\mathbf{f}(\mathbf{v}_1)-\mathbf{f}(\mathbf{v}_2)\right\|\le L_{f} \left\|\mathbf{v}_1-\mathbf{v}_2\right\|$, with $L_f=36000$ being the Lipschitz constant.
This quantity is excessively large, and using the classic approach would lead to very conservative results \cite{zemouche2013}. 
Therefore, we consider the less restrictive QUAD condition instead.
Indeed, the function $\mathbf{f}$ satisfies the QUAD condition by setting $\rho=L_f$ and $\mathbf{P}=\mathbf{I}_N$ \cite{delellis2011}.
%
%
However, to find the lowest value of $\rho$, we need to find the optimal matrices $\mathbf{P}$ and $\mathbf{L}$ such that condition (i) and (ii) are fulfilled.
To do so, we rewrite these two conditions as a constrained nonlinear multivariate optimization problem, that is, $\min_{\mathbf{v}_1,\mathbf{v}_2,\mathbf{P},\mathbf{L}}\{\rho\}$ s.t. (i) and (ii) hold. We solve it using the Matlab's Optimization Toolbox, and we find that $\rho=27.86$, 
$$
\mathbf{P} = \left[ {\begin{array}{ccc}
   {0.6995} & {0} & { -0.004}\\
   {0} & {0.001} & {0}\\
   {-0.004} & {0} & {3}
  \end{array} } \right], \mathbf{L} = \left[ {\begin{array}{ccc}
   {21} & {-141}\\
   {0.12} & {14705}\\
   {267} & {-0.18}
  \end{array} } \right],
$$
%
%
%
%
%
%
%
and $\mathbf{Q}=\mbox{diag}\{29.437, 29.437, 1624.266\}$. 
Note that both $\mathbf{P}$ and $\mathbf{Q}$ are symmetric and positive definite. 
Moreover, $\lambda_{\R{min}}(\mathbf{Q}) = 29.437$; hence, the third condition (iii) $\lambda_{\R{min}}(\mathbf{Q})>27.8644$ of Proposition \ref{propo:1} is fulfilled and the synthesis of the observer is complete.

\subsection{Zero Average Dynamics (ZAD)}
The ZAD controller is a quasi-sliding technique, where the goal is forcing the switching function to be zero on average over a finite period of time.
This strategy was originally developed for controlling DC-DC power converters in \cite{fosas2000}; however, it has been also recently used for controlling gene expression in synthetic biology \cite{fiore2015}.
This strategy has been shown to provide low regulation error and most importantly fixed switching frequency.
In fact, when compared with traditional sliding control, where there is typically an infinite number of commutations (thus inducing chattering), ZAD control guarantees a finite number of switches over a finite period of time. 
Specifically, we consider the control input $u(t)$ to be given by a centered PWM of the form
\begin{equation}
u(t) = \begin{dcases}
{\mu}, &  {\textup{if } kT\le t\le kT+d_k/2} \\
{-\mu}, & {\textup{if } kT+d_k/2<t<(k+1)T-d_k/2}\\
{\mu}, &  {\textup{if } (k+1)T-d_k/2\le t<(k+1)T}
\end{dcases},
\label{eq:ZAD:control}
\end{equation}
where $T$ is the switching period, $k\in\{0,1,2,\dots,m\}$, with $m$ being the number of samples, $\mu>0$, and $d_k$ is the \textit{duty cycle} (i.e.~the time that the switch remains ON).
We choose the sliding surface to be a combination of a proportional and a derivative term, i.e.~$S(\psi(t), \dot{\psi}(t)) := \psi(t)-k_{\mathrm{s}}\dot{\psi}(t)$, with $k_{\mathrm{s}}>0$ \cite{angulo2005}.
Then, the main design problem is to find the duty cycle $d_k$ such that
\begin{equation}
\int\limits_{kT}^{(k+1)T}{S(\psi, \dot{\psi}) \ \R{d}t} = 0, \quad \forall\quad k\in\{1,2,\dots,m\}.
\end{equation}
As pointed out in \cite{angulo2005}, solving this transcendental equation for each time interval demands high computation cost.
Therefore, using a linear approximation of the sliding surface and solving for the variable $d_k$ yields a simple expression for the duty cycle and its normalization $d_{\mathrm{c}}:=d_k/T$ (for further details about this approximation, see \cite{angulo2005} and references therein):
\begin{equation}
d_k=\frac{2S(kT)+T\dot{S}_2(kT)}{\dot{S}_2(kT)-\dot{S}_1(kT)},
\label{eq:zad}
\end{equation}
where $\dot{S}_1(kT)$ and $\dot{S}_2(kT)$ are given by
\begin{align}
\dot{S}_1(kT) &= \dot{S}(kT)\rvert_{u=\mu} = (\dot{\psi}(kT)+k_{\mathrm{s}}\dot{\psi}(kT))\rvert_{u=\mu},\\
\dot{S}_2(kT) &= \dot{S}(kT)\rvert_{u=-\mu} = (\dot{\psi}(kT)+k_{\mathrm{s}}\ddot{\psi}(kT))\rvert_{u=-\mu}.
\end{align}
Note that, in order to find $\dot{S}_1(kT)$ and $\dot{S}_2(kT)$, we need the knowledge of $\alpha(kT)$; instead, we will use the estimation $\hat{\alpha}(kT)$ provided by the observer.
\subsection{Minimal Control Synthesis (MCS)}
Minimal Control Synthesis  (MCS) is a strategy used to determine the control gains of a classical Model-Reference Adaptive Controller (MRAC) \cite{stoten1990}.
The aim is that of making the plant, typically assumed to be in controllable canonical form, track asymptotically the output, $\mathbf{y}_\mathrm{m}$, of some linear reference model described by the matrices $\mathbf{A}_\mathrm{m}$, $\mathbf{B}_\mathrm{m}$, $\mathbf{C}_\mathrm{m}$, so as to make the error $\mathbf{e}_\mathrm{m}:=\mathbf{y}_\mathrm{m} - \mathbf{y}$ asymptotically null.
To this aim, the control input is selected as $u(t) = -\B{K}(t)\B{x}(t)$,
%
%
where 
\begin{equation}\label{eq:K_x_MCS}
\B{K} (t) = k_\R{P} \left( w(t) \B{x}\T(t) \right) + k_\R{I} \left(\int_{0}^{t} w(\tau) \B{x}\T(\tau) \R{d} \tau \right),
\end{equation}
with $k_\R{P}$ and $k_\R{I}$ being constants, $w(t) := \B{B}_\R{m}\T \B{P} \B{e}_\R{m}(t)$ and $\B{P}$ being a symmetric positive definite matrix that verifies the Lyapunov equation $\B{P} \B{A}_\R{m} + \B{A}_\R{m}\T \B{P} = - \B{Q}$, with $\B{Q}>0$.
Here, we present an empirical implementation of the MCS on the NLG model in equations \eqref{main:eq} by selecting $\B{A}_\R{m} = \B{A}$, $\B{B}_\R{m} = \B{B}$, $\B{C}_\R{m} = \B{C}$, $k_\R{P} = 10^3$, $k_\R{I} = 10^4$, and $\B{P} = \B{I}_3$.
Note that all nonlinear terms in model \eqref{main:eq} are assumed as nonlinear disturbances acting on the linear terms of the plant. 
As in the case of the ZAD controller, again here the observer is needed to estimate the whole state vector required to compute the control action $u(t)$.
\section{Numerical Results}
\label{sec:Numerical Results}
In this section, we test the two control strategies described in the previous one.
In particular, we consider two different tests that have been widely used in the literature for evaluating the performance of shimmy control techniques \cite{pouly2011}.%
\begin{figure}[t]
\centering {
\subfigure[]
{\label{fig:4:a}
{\includegraphics[scale=0.2]{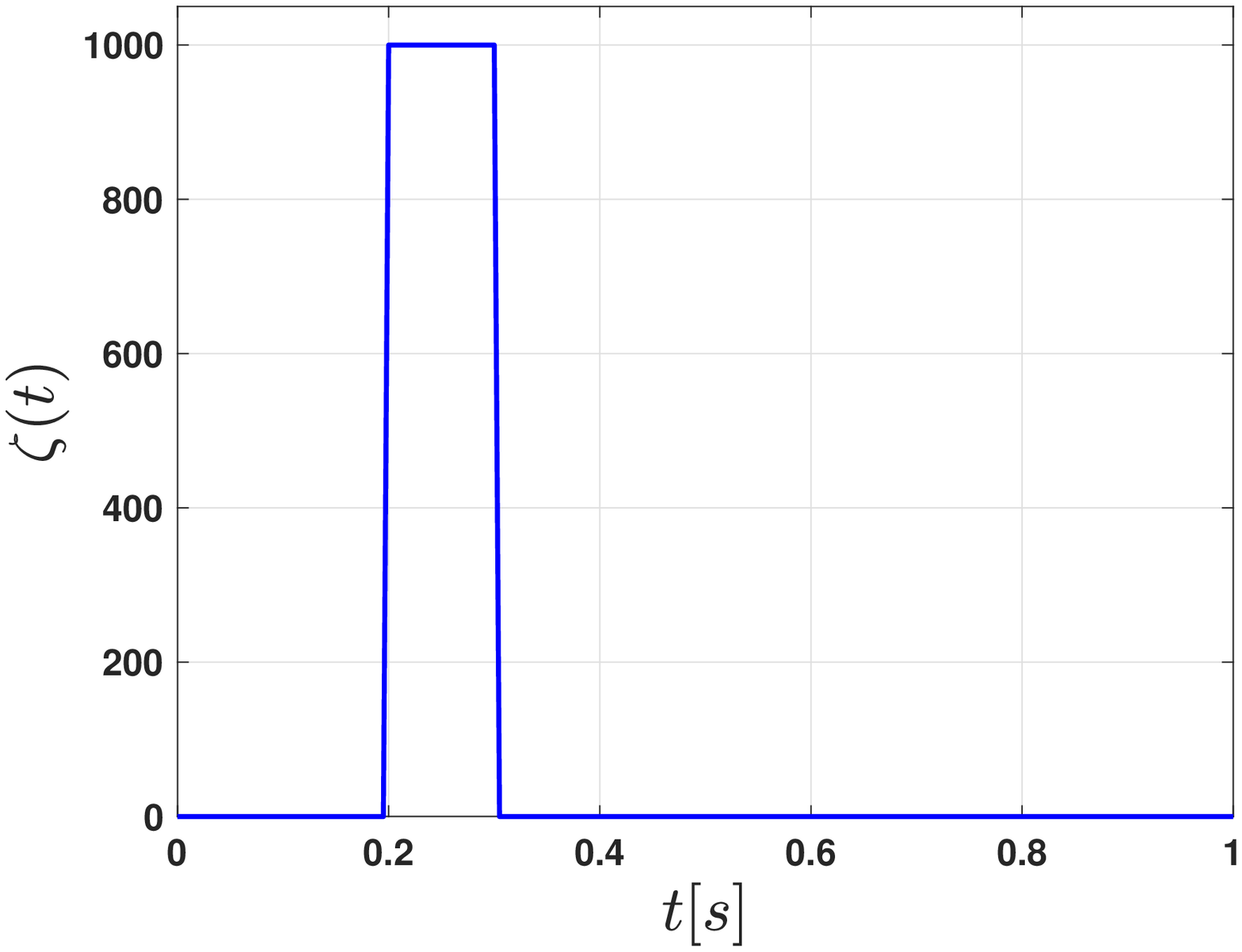}}}
\subfigure[]
{\label{fig:4:b}
{\includegraphics[scale=0.2]{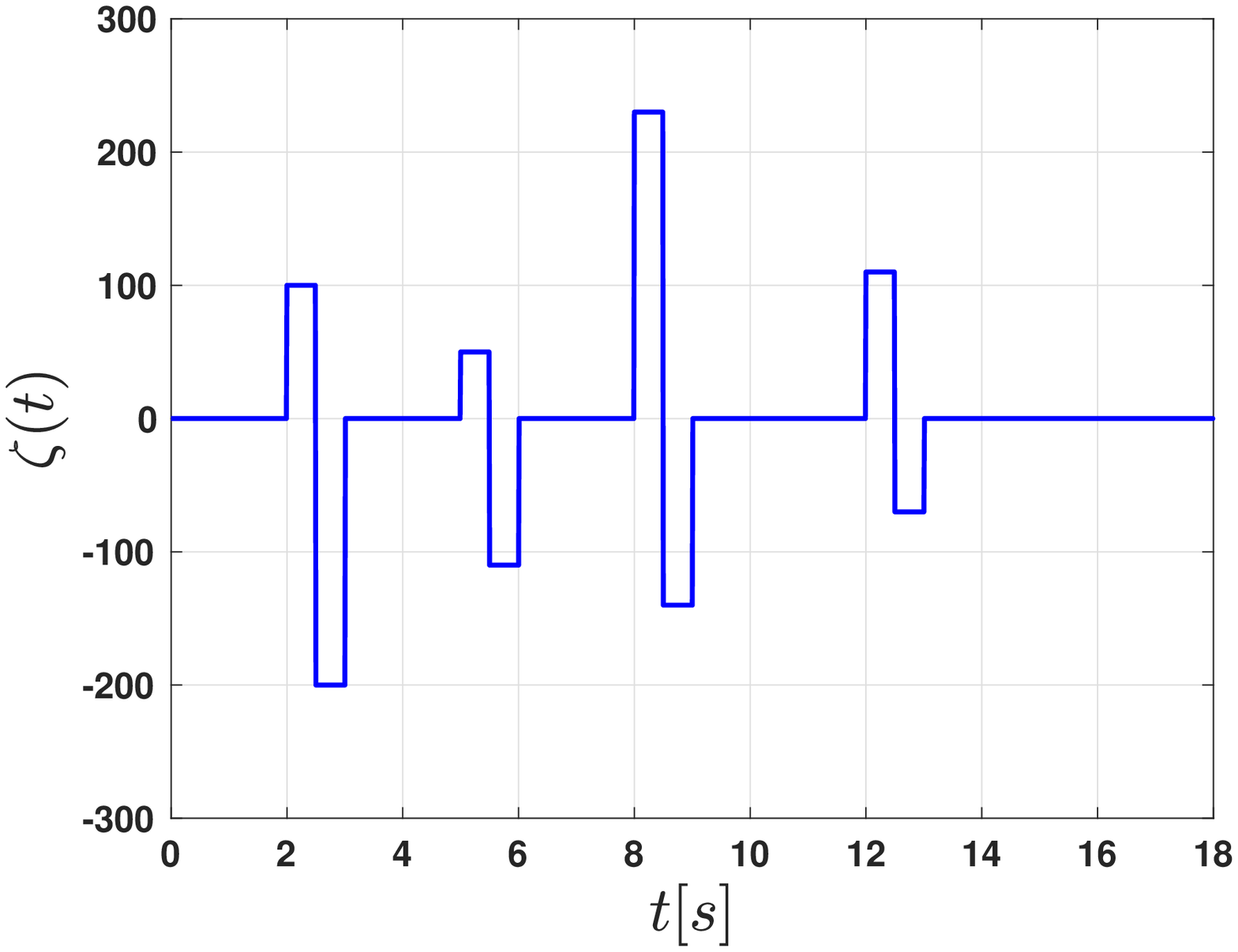}}}
\subfigure[]
{\label{fig:4:c}
{\includegraphics[scale=0.2]{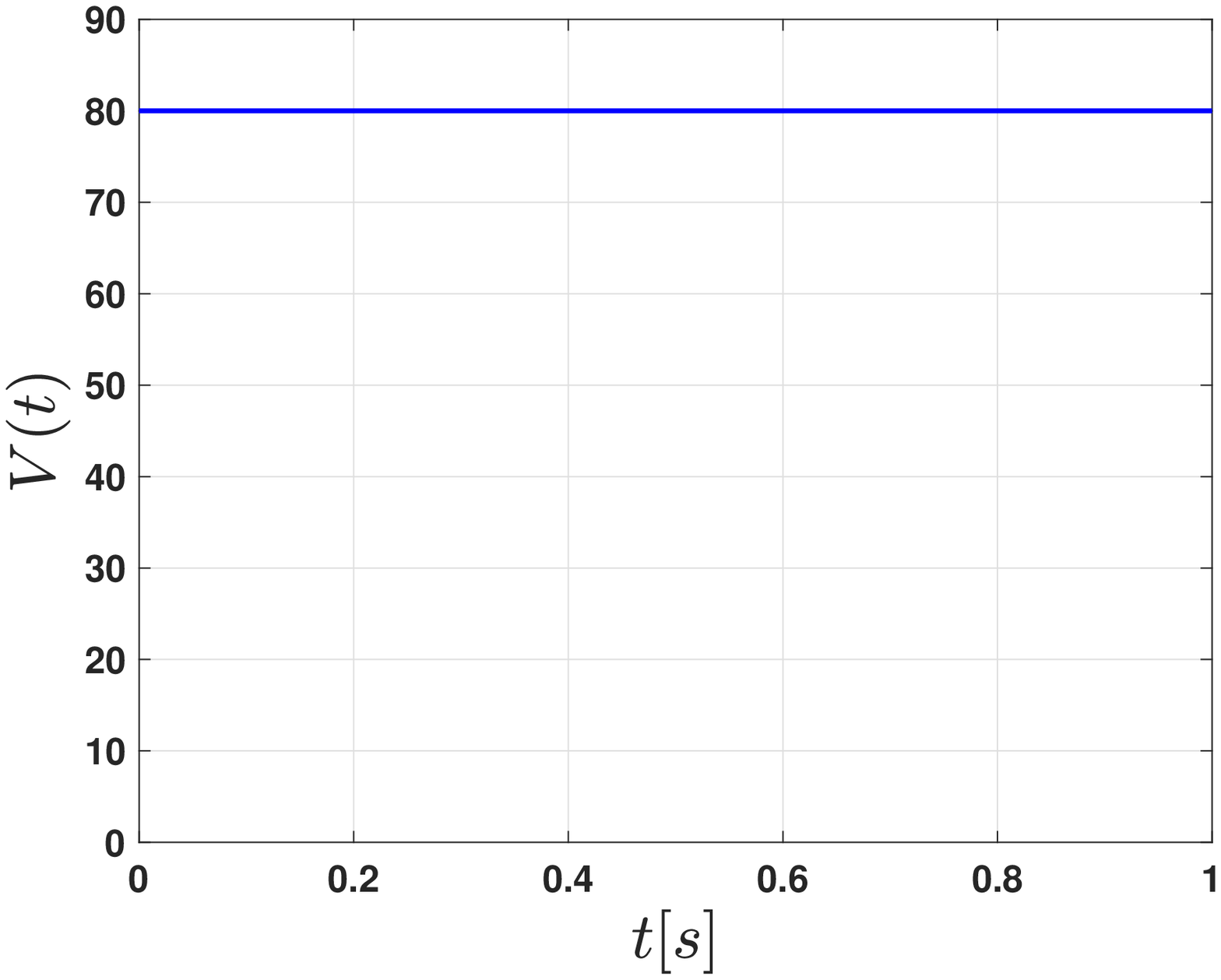}}}
\subfigure[]
{\label{fig:4:d}
{\includegraphics[scale=0.2]{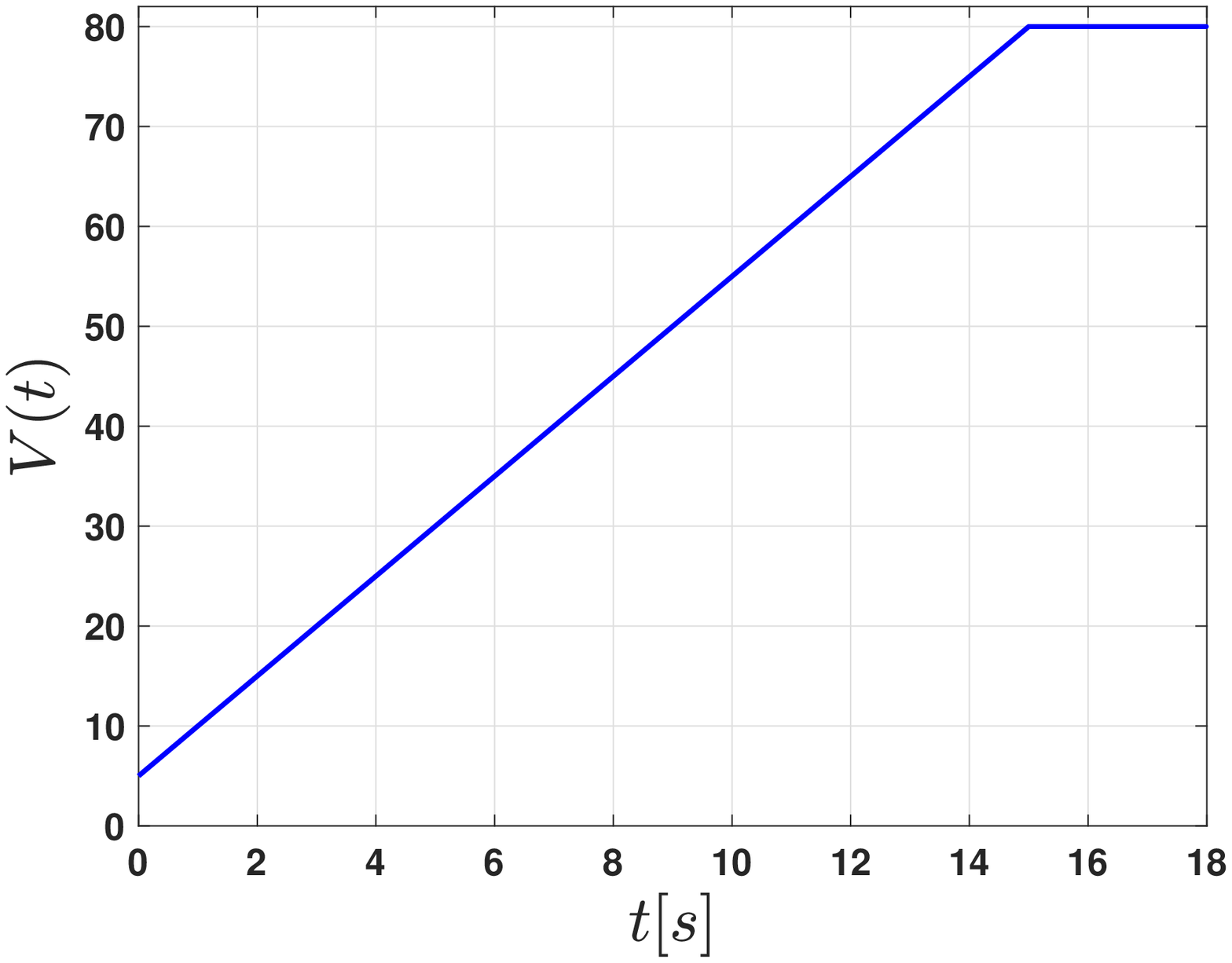}}}
\subfigure[]
{\label{fig:4:e}
{\includegraphics[scale=0.2]{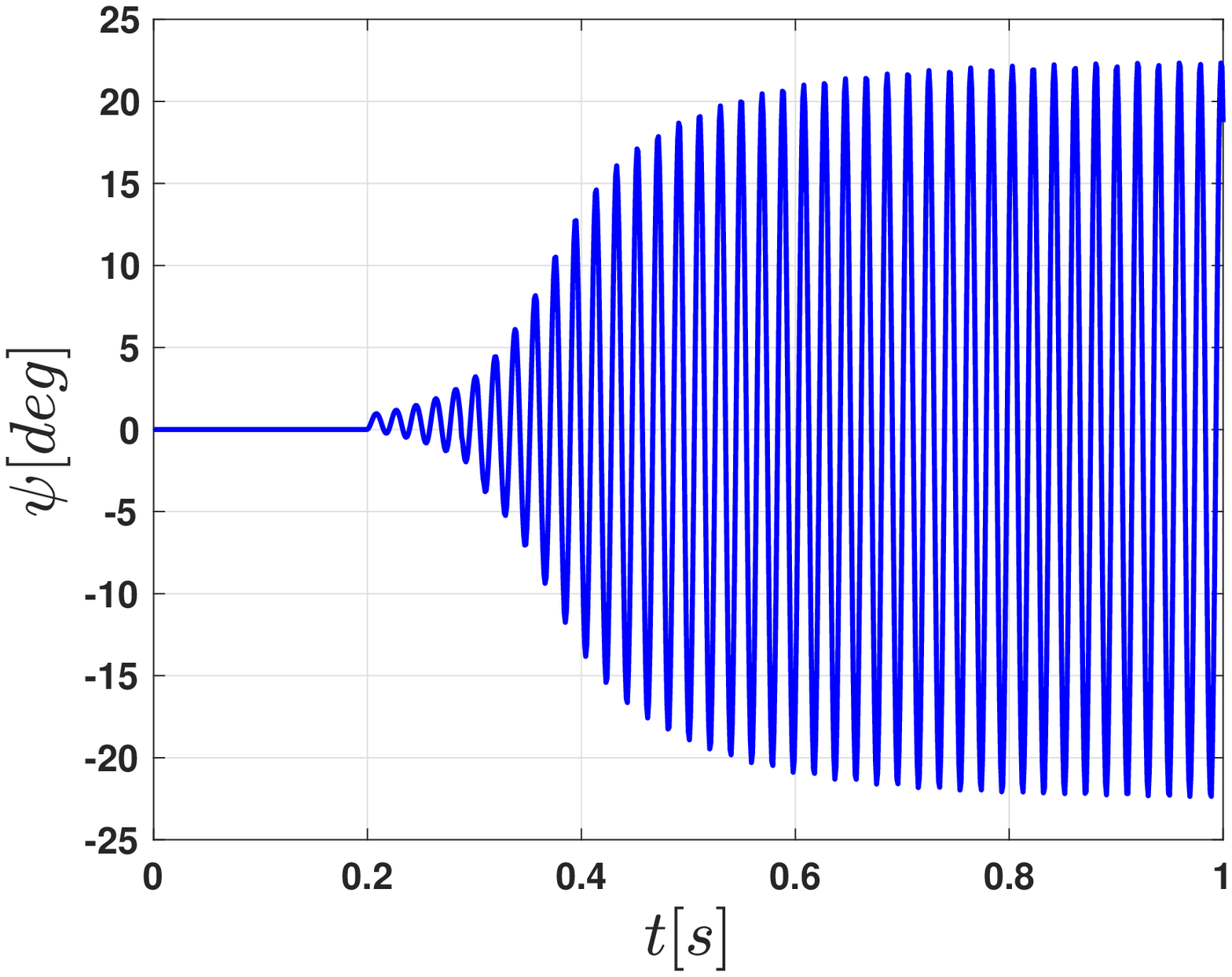}}}
\subfigure[]
{\label{fig:4:f}
{\includegraphics[scale=0.2]{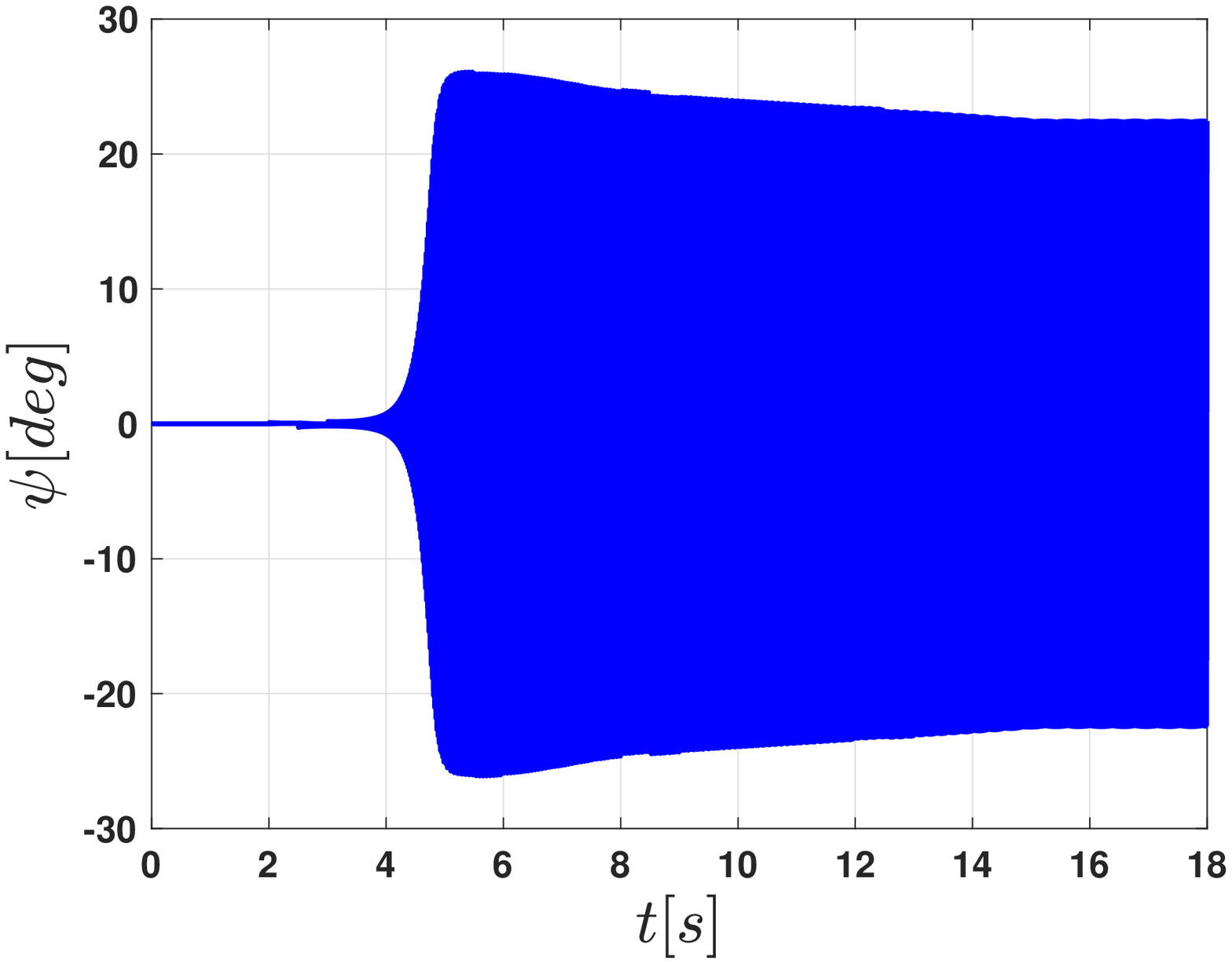}}}
}
\caption{Time trajectories of disturbance $\zeta$ (a, b), velocity $V$ (c, d) and wheel angle $\psi$ (e, f) in the cases of Test 1 (a, c, e) and Test 2 (b, d, f).}
\label{fig:4}
\end{figure}%
\begin{itemize}
    \item \textit{Test 1: Tire damage.} In this case, we assume a constant speed $V=80$ with zero initial conditions.
Then, for $0.2\le t\le0.3$, an impulse function acting as a disturbance torque in \eqref{main:eq} is present, that is, $\zeta(t) = 1000 \ \R{nM}$, as shown in Figure \ref{fig:4:a}.
\item \textit{Test 2: Taxiing on non-uniform road.} In this scenario, the aircraft is taxiing with increasing velocity, that is the velocity varies according to the ramp depicted in Figure \ref{fig:4:d}, while the disturbance $\zeta(t)$ is given by the function shown in Figure \ref{fig:4:b}, which simulates potholes on the road.
Note that the velocity range adopted here is similar to that explored in the bifurcation diagrams of Figure \ref{fig:1}, where it has been shown that shimmy occurs for velocities grater than 20 m/s. 
\end{itemize}
To further test the robustness of the control strategies, we add uncertainty to the nonlinear function $\mathbf{f}(\mathbf{x})$ by using the piece-wise smooth approximation \eqref{eq:M_nonsmooth}-\eqref{eq:F_nonsmooth} for the NLG model, while the smooth approximation \eqref{eq:M_smooth}-\eqref{eq:F_smooth} is used for the observer and the controllers. %
For the sake of comparison, we first present the open-loop responses of the NLG in Tests 1 and 2, in Figures \ref{fig:4:e} and \ref{fig:4:f}, respectively.
Note that in both cases oscillations with large amplitude are present.
\subsection{Test 1: Tire damage}
\begin{figure}[t]
\centering {
\subfigure[]
{\label{fig:5:a}
{\includegraphics[height=5cm]{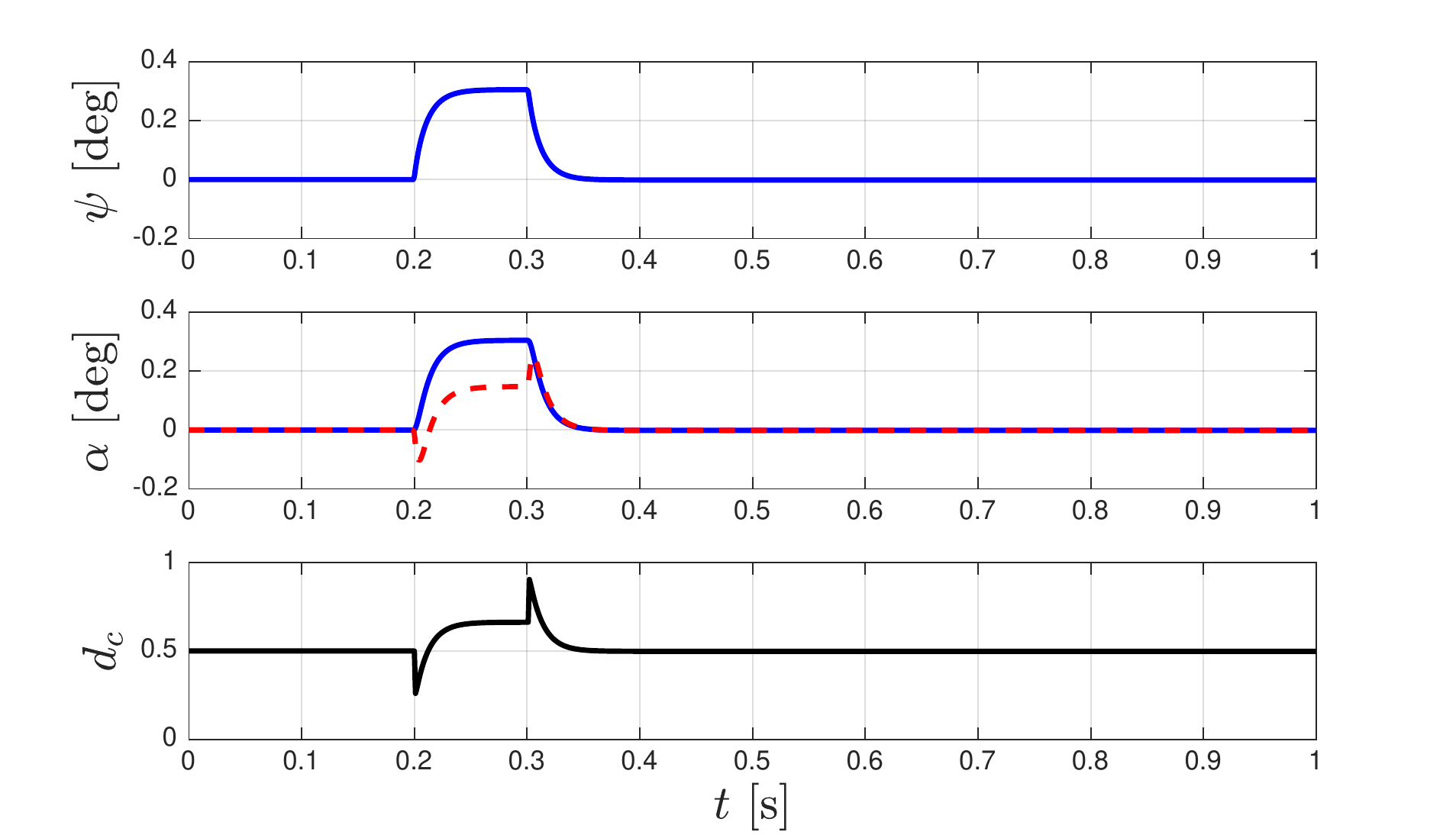}}}
}
\subfigure[]
{\includegraphics[height=5cm]{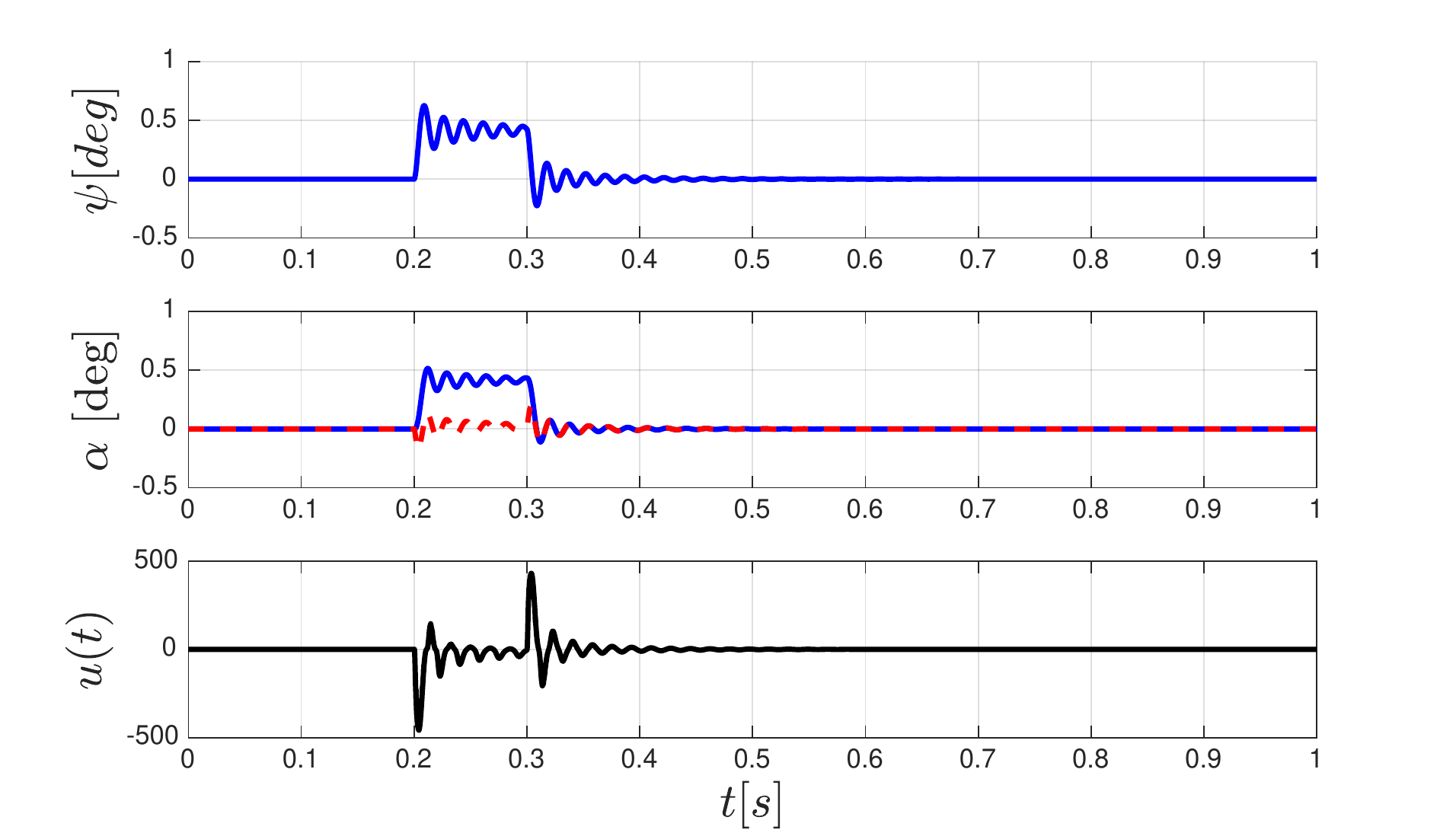}
\label{fig:5:b}}
\caption{Time response of the NLG under impulsive disturbance, controlled by (a) ZAD strategy and (b) MCS. In the top, middle and bottom panels, $\psi(t)$, $\alpha(t)$, and $d_{\mathrm{c}}(t)$ (or $u(t)$) are shown respectively. The red dashed line represents the estimation $\hat{\alpha}(t)$ made by the observer, whereas the solid line line is $\alpha(t)$.}
\label{fig:5s:}
\end{figure}

We first test the ZAD controller in the case of constant speed and impulsive disturbance.
In particular, we set the sampling period $T=10^{-3}$, the constant $k_{\mathrm{s}} = 0.5$, the control gain $\mu=1000$ and zero initial conditions.
The time response of the closed-loop systems is shown in Figure \ref{fig:5:a}, where it is evident that the yaw angle $\psi$ is driven to zero and the maximum overshoot is less than one degree.
Furthermore, the normalized duty cycle $d_{\R{c}}$ converges to a constant value and hence the control action exhibits fixed switching frequency.
%
%
%

Next, we perform Test 1 for the NLG controlled by MCS.
The time trajectories are shown in Figure \ref{fig:5:b}, which shows that the controller is able to suppress the undesired oscillations despite the presence of uncertain nonlinear terms, while guaranteeing boundedness of the adaptive control action.
%
%
%
%
\subsection{Test 2: Taxiing on non-uniform road}
\begin{figure}[t]
\centering {
\subfigure[]
{\label{fig:6:a}
\includegraphics[height=5cm]{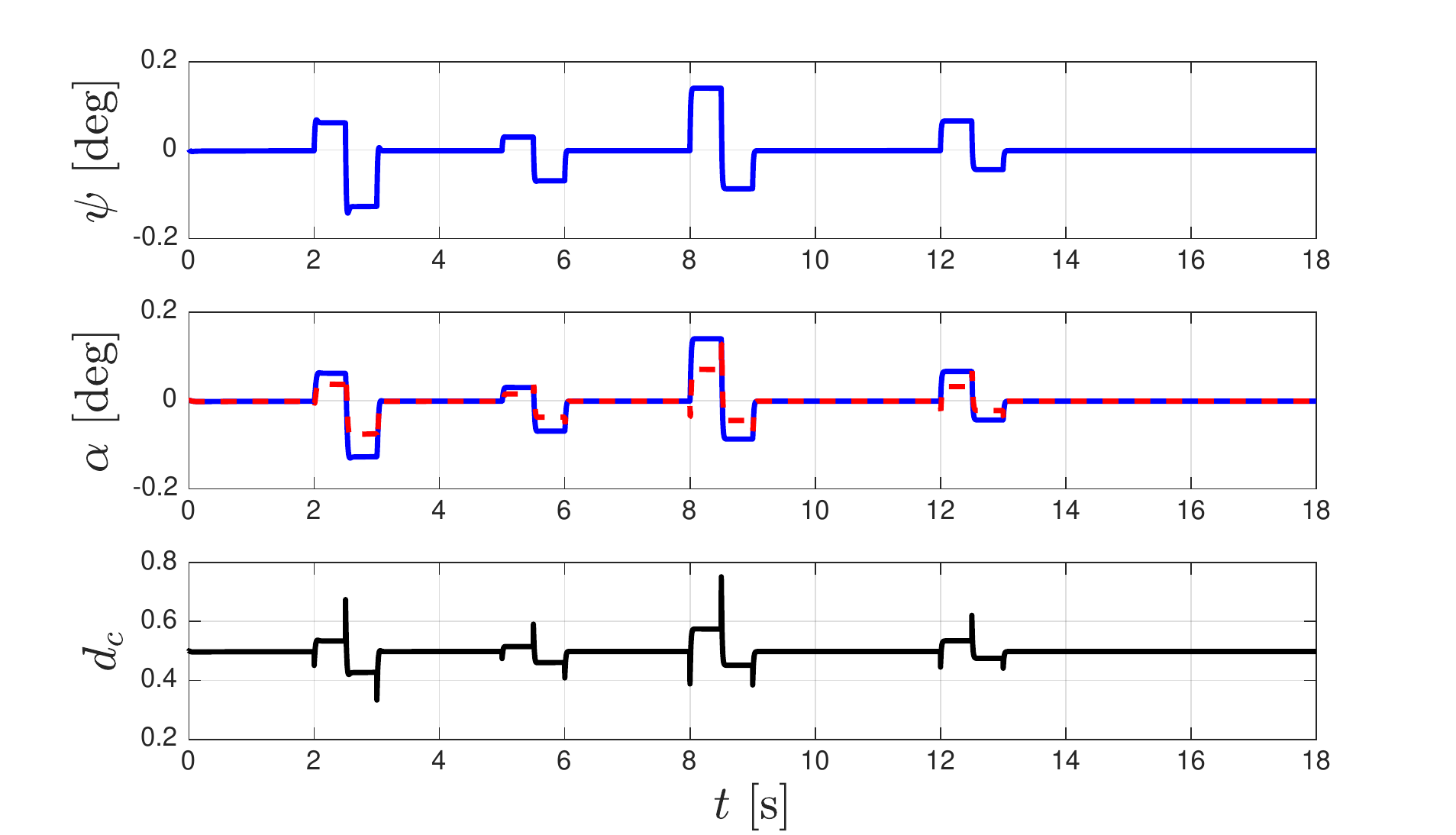}}
}
\subfigure[]
{\label{fig:6:b}
\includegraphics[height=5cm]{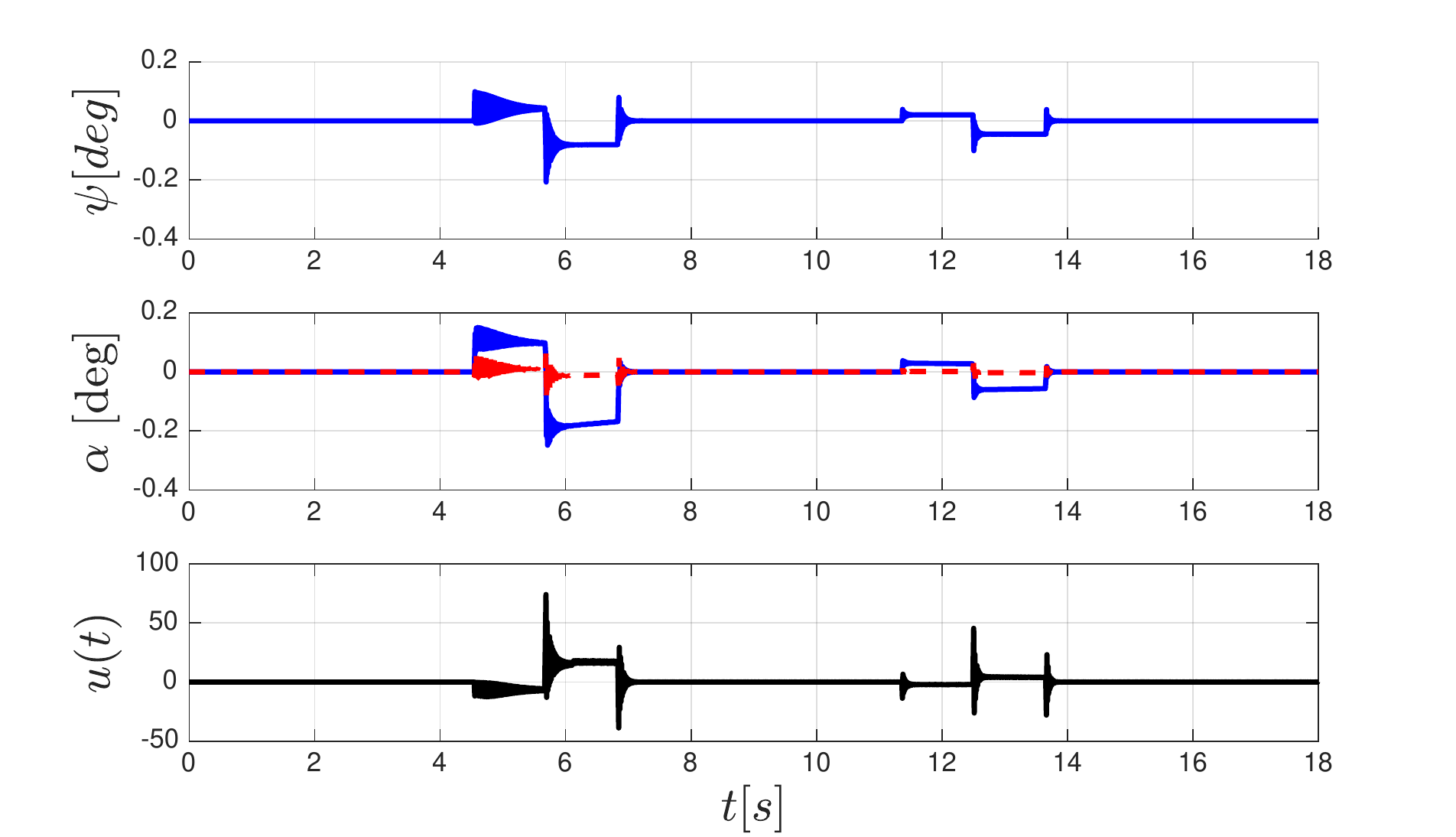}}
\caption{Time response of the NLG for time-varying velocity and non-uniform road, controlled by (a) ZAD strategy and (b) MCS. In the top, middle and bottom panels, $\psi(t)$, $\alpha(t)$, and $d_{\mathrm{c}}(t)$ (or $u(t)$) are shown respectively.}
\label{fig:9s}
\end{figure}

The time response of the NLG controlled by ZAD and MCS is shown in Figures \ref{fig:6:a} and \ref{fig:6:b}, respectively.
Note that even in this case, where there are multiple perturbations (due to the potholes), the controllers are able to effectively suppress shimmy.
%
%
%
%
%

%
%
%
\section{Conclusions}
\label{sec:Conclusions}
We have proposed the use of ZAD and MCS control approaches for suppressing shimmy in a NLG with uncertain non-linearities and partial state measurements.
In so doing, we adopted less conservative conditions for designing observers with nonlinear terms, so that large overshoots can be avoided.
Using numerical simulations, we showed the effectiveness of the proposed control strategies under two representative scenarios.
Future work is needed to develop stability analysis of the closed-loop systems, together with an accurate performance assessment of both control techniques.
In addition, the control approaches can be also tested using more realistic NLG models as those considered in \cite{howcroft2015}.
\bibliographystyle{IEEEtran}
\bibliography{Refer_I}
\end{document}